\documentclass[a4paper,11pt]{article}
\usepackage{jheppub} 

\usepackage{color}
\usepackage{graphicx}
\usepackage{textcomp}
\usepackage{amsmath,amssymb}
\usepackage{enumerate}
\usepackage[normalem]{ulem}
\newcommand{\mathsym}[1]{{}}

\newcommand{\ba}{\begin{array}}
\newcommand{\ea}{\end{array}}
\newcommand{\be}{\begin{equation}}
\newcommand{\ee}{\end{equation}}
\newcommand{\beqa}{\begin{eqnarray}}
\newcommand{\eeqa}{\end{eqnarray}}
\def\321{$SU(3)\times SU(2)\times U(1)$}
\def\mt{$\mu$-$\tau$~}

\newcommand{\Upmns}{U_{\rm PMNS}}
\newcommand{\Dms}{\Delta m^2_{\rm sol}}
\newcommand{\Dma}{\Delta m^2_{\rm atm}}
\newcommand{\group}[1]{\textlbrackdbl #1\textrbrackdbl}
\newcommand{\state}[1]{\textbar #1\textrangle}
\def\gm{\gamma}
\def\vt{\vartheta}
\def\et{\eta}
\def\ps{\psi}
\def\ph{\phi}

\title{A massless neutrino and lepton mixing patterns from  finite discrete subgroups of
$U(3)$}

\author[a]{Anjan S. Joshipura,}
\author[b]{Ketan M. Patel}

\affiliation[a]{Physical Research Laboratory, Navarangpura, Ahmedabad 380 009, India.}
\affiliation[b]{Istituto Nazionale Fisica Nucleare, Sezione di Padova, I-35131 Padova,
Italy.}

\emailAdd{anjan@prl.res.in}
\emailAdd{ketan.patel@pd.infn.it}

\abstract{Finite discrete subgroups of $U(3)$ as possible flavour symmetries $G_f$ for a massless neutrino
with predictive mixing angles are studied. This is done by assuming that
a residual symmetry
$S_\nu$ appropriate for  describing  a massless neutrino is contained in
$G_f$. It is shown that all the
groups $G_f$ admitting  three dimensional faithful irreducible representation and  generated from a
specific set of $3\times 3$ matrices imply only one of the three  flavour compositions for the
massless state namely, unmixed, maximally mixed  with equal probabilities and bimaximally mixed
with probabilities $(0,1/2,1/2)$ and their permutations. This result holds irrespective of the order
of $G_f$ and the choice of $S_{\nu}$ within it. All of these lead to unfavorable leading order
prediction for the solar mixing angle. Neutrino mixing pattern is then numerically investigated in
case of subgroups of $U(3)$ with order less than 512 and it is found that only one of these can lead
to a massless neutrino and leading order predictions for all the  mixing angles close to their
experimental values. Ways to correct for the solar angle prediction are proposed and two concrete
examples giving  the observed mixing pattern are discussed.}

\begin{document} 
\maketitle
\flushbottom

\section{Introduction}
The orderly pattern found in leptonic mixing may be signaling the existence of some underlying
flavour symmetry, see \cite{reviews, Ishimori:2010au} for
some recent reviews. The mixing pattern is quite well-known by now but it is still difficult to fix
a discrete symmetry responsible for it. A systematic approach to look for such symmetries has been
pursued vigorously \cite{symmetry}. Basic idea in this approach is to (a) first obtain residual
symmetries $G_\nu$ and $G_l$ of the neutrino and the charged lepton mass matrices based on the
observed mixing pattern and (b) look for the flavour symmetry group $G_f$ which contain $G_\nu$,
$G_l$ as subgroups. $G_\nu$ is usually a $Z_2\times Z_2$ symmetry in case of three massive Majorana
neutrinos with unconstrained masses and $G_l$ is some discrete subgroup of $U(1)\times U(1) \times
U(1)$. The advantage of this approach is that mixing pattern is completely fixed by the choice of
$G_\nu$ and $G_l$ without any detailed knowledge of the underlying theory. Moreover, a general
prescription can be formulated  \cite{Lam:2011ag}  which leads to the desired mixing
from some underlying Lagrangian invariant under $G_f$ spontaneously broken to $G_\nu$ and $G_l$.
This basic framework has been used \cite{reviews, Ishimori:2010au, symmetry}
to obtain zeroth order mixing patterns such as tri-bimaximal (TBM) and bimaximal (BM) and also to
obtain some reasonable predictions on the mixing angles but flavour groups which can predict all the
mixing angles in leading order within their $2\sigma$ values are very few and large with order
$>600$ \cite{Holthausen:2012wt}.

The above framework does not put any restrictions on neutrino masses. The present neutrino data are
still consistent with  the quasi degenerate neutrinos or a spectrum with a massless neutrino both in
case of the normal and inverted mass hierarchies. It would be appropriate to look for the
modification of the above scenario which yields either of these spectrum. The case of the quasi
degenerate neutrino was considered in \cite{Hernandez:2013vya} while the flavour symmetries
appropriate for obtaining a massless state were discussed in \cite{Joshipura:2013pga}. Both, the
choices for the residual symmetries and the flavour groups become different in theses cases. If
neutrinos are Majorana particles and one of them is massless then one can redefine the
phase of the massless field. Thus, one of the $Z_2$ symmetries of the neutrino mass matrix used as a
member of $G_\nu$ in the usual approach gets replaced by a $Z_n$ symmetry with the corresponding
$3\times 3$ matrix $S_\nu$ having ${\rm Det.}(S_\nu)\not =\pm 1$. As a result, the group embedding
this also needs to be a discrete subgroup (DSG) of $U(3)$ rather than that of $SU(3)$. This idea was
elaborated in \cite{Joshipura:2013pga} and the group series $\Sigma(2n^2)$
\cite{Ishimori:2010au} and $\Sigma(3n^3)$ as well as the group $S_4(2)$ \cite{Ludl:2010bj} were
studied as possible flavour symmetries of the massless neutrino. The groups $\Sigma(2n^2)$ were
shown to lead to a neutrino mass matrix with \mt symmetry for arbitrary $n$. The groups
$\Sigma(3n^3)$ were shown to lead to democratic mixing while $S_4(2)$ provided an example of
symmetry leading to the BM mixing pattern. None of these groups could predict all three mixing
angles correctly but both $S_4(2)$ and $\Sigma(2n^2)$ could yield a good zeroth order
predictions and perturbations leading to the correct mixing patterns were also studied in case of
$S_4(2)$.

Aim of the present paper is to make an exhaustive search for an appropriate flavour group and see 
if a massless neutrino and realistic mixing angles can be obtained with some flavour symmetry. This
search leads to a ``no-go result'' which demonstrates that the solar angle cannot be predicted
correctly at the leading order in a very large class of groups to be
specified as we go along. In all these groups, the
flavour composition of the massless state gets determined irrespective of the nature of the group
and for all possible choices of $G_\nu$ and $G_l$ within it. The allowed compositions are only three
types. The trimaximal  with probability to be in different flavour states as $\frac{1}{3}(1,1,1)$ 
or bimaximal with probabilities  $\frac{1}{2}(0,1,1)$ and its permutations and  trivial one
$(1,0,0)$ and its permutations. This immediately results in wrong prediction for the solar angle in
case of  the normal hierarchy. In the case of inverted hierarchy, bimaximally mixed massless state
provides a good leading order approximations to the reactor and atmospheric mixing angles but the
solar angle turns out to be maximal as we will show later. This feature follows in a large class of
groups which are DSG of $U(3)$ having a three dimensional faithful irreducible representation (IR).
In particular, this result holds for all but one group having order $<512$.

This paper is organized as the follows. We briefly review the idea of predicting a massless
neutrino within the realm of the discrete symmetries in the next section. In section
\ref{analytical}, we provide a simple analytic study of the large class of DSG of $U(3)$ and
discuss the allowed  mixing patterns with a massless neutrino. The predictions of lepton mixing
angles obtained by an exhaustive numerical scan of small DSG of $U(3)$ are presented in section
\ref{scan}. We discuss in section \ref{correction} some alternatives which can lead to correct solar
mixing angles as well and discuss specific scenario capable of reproducing all  three mixing
angles in  the experimentally allowed range. Finally, we summarize our study in section
\ref{summary}.

\section{Massless neutrino and flavour symmetries}
\label{review}
We first review a framework  relating flavour symmetries to neutrino mixing and its generalization
which can also yield a massless neutrino. Assume that the neutrino (charged lepton) mass matrix in
arbitrary basis is invariant under a set of unitary generators $S_i$ ($T_\alpha$) which form a
discrete group $G_\nu$ ($G_l$). All $S_i$ are assumed to commute and are diagonalized by a matrix
$V_\nu$. Similarly, the commuting set $T_\alpha$ is diagonalized by $V_l$. The invariance of the
leptonic mass matrices under the respective groups can be used to show that the neutrino mixing
matrix $\Upmns$ is determined in terms of $V_\nu$ and $V_l$ \cite{symmetry}:
\be 
\Upmns=P_l V_l^\dagger V_\nu P_\nu ~,\ee
where $P_l$ and $P_\nu$ are arbitrary diagonal phase matrices.

The minimal group $G_\nu$ can be constructed from the observation that mass terms for the Majorana
neutrinos remain unchanged under a  change of sign of any of the neutrino mass eigenstates. Thus
neutrino mass matrix in the mass basis is trivially invariant under
\be \label{snu0}
s_1={\rm Diag.}(1,-1,-1)~,~~s_2={\rm Diag.}(-1,1,-1)~~{\rm and}~~s_3=s_1 s_2~, \ee
where Det$(s_{i})$ is chosen +1. Any two of these define a $Z_2\times Z_2$ symmetry. This implies
that 
\be \label{si}
S_i=V_\nu s_i V_\nu^\dagger~ \ee
correspond to $G_\nu =Z_2\times Z_2$ symmetry in a  basis labeled by $V_\nu$. This along with the 
\be \label{talpha}
T_l=V_l~{\rm Diag.}(e^{i \phi_e},e^{i \phi_\mu},e^{i \phi_\tau})~V_l^\dagger, \ee
with $\phi_{e,\mu,\tau}$ being some discrete phases can be used as residual symmetries to be
embedded in a bigger group $G_f$.

When one of the neutrinos is massless, then phase of the corresponding field can be changed without
affecting neutrino mass term. In this case, one of the $S_i$ say $S_1$ can be replaced by
\cite{Joshipura:2013pga} 
\be \label{snu1}
S_{1\nu}= V_\nu ~{\rm Diag.}(\eta,1,-1)~ V_\nu^\dagger ~,\ee
with $\eta^N=1$ and $N\geq 3$. The $S_{1\nu}$ forms a $Z_N$ ($Z_{2 N}$) group for even (odd) $N$.
This along with $S_{2,3}$ provide  possible  residual symmetries if one
neutrino is massless.
We are assuming here that the two massive neutrinos are non-degenerate.
$S_\nu$ has to be chosen differently if this not the case
\cite{Hernandez:2013vya}. We shall not entertain this possibility here.

Note that, all the eigenvalues of $S_{1\nu}$ as defined above are different. This uniquely fixes
all the mixing angles in $V_\nu$ and it is sufficient to take this  as a residual symmetry of the
neutrino mass matrix and look for embedding of this $G_\nu$ and $G_l$ into a bigger group $G_f$.
Arbitrariness in one of the mixing angles would remain if $S_{1\nu}$ above is replaced by
\be \label{zn}
S_{1\nu}'= V_\nu ~{\rm Diag.}(\eta,1,1)~V_\nu^\dagger ~.\ee
We shall use either $S_{1\nu}$ or $S_{1\nu}'$ as possible $Z_N$ in the following. As for $G_l$, we
shall choose $T_l$ defined in Eq. (\ref{talpha}) which uniquely fixes $V_l$ if all phases are
different. We then look for the discrete subgroups of $U(3)$ which contain $T_l$ and $S_{1\nu}$ or
$S_{1\nu}^\prime$ as elements. A complete classification of all discrete subgroups of $U(3)$ is not
available but our analysis would encompass all the known DSG of $U(3)$ with a three dimensional
faithful irreducible representation and many having (2+1)-dimensional reducible representation
\cite{Parattu:2010cy, Ludl:2010bj}.

\section{Massless neutrino and lepton mixing from discrete subgroups of $U(3)$}
\label{analytical}
Let us first review here some available information on the DSG of $U(3)$ which will be used in our
analysis. Details can be found in \cite{GrimusLudl:groups} and specially in
\cite{Ludl:2010bj}. All the DSG of $U(3)$ which are also the subgroups of $SU(3)$ have been
classified. The complete list can be found for example in 
\cite{GrimusLudl:groups, Grimus:SU3}. They are characterized in terms of some small set of
$3\times 3$ matrices  which are used to generate elements of various DSG of
$SU(3)$. The present knowledge on
DSG of $U(3)$ which are not the subgroups of $SU(3)$ is partial. We will concentrate in this section
on such subgroups of $U(3)$ admitting a faithful three dimensional irreducible representation. It is
convenient to divide these groups in two categories:
\begin{itemize}
\item[(X)] Those which are generated by a specific set of generators $R,~S,~T,~U,~V$ and $W$ as given in
\cite{Ludl:2010bj} and reproduced here in Table \ref{generators} and 
\item[(Y)] those which are generated by more special textures labeled as $X_1,...,X_{10}$ in Table
\ref{generators}.
\end{itemize}

\begin{table}[!ht]
 \begin{center}
 \begin{footnotesize}
 \begin{math}
 \begin{tabular}{ccc}
 \hline
 \hline
 \multicolumn{3}{c}{Generators of finite DSG of $U(3)$ of category (X)}\\
 \hline
$S(n,a,b,c)=\left(\ba{ccc}
\et^a&0&0\\
0&0&\et^b\\
0&\et^c&0\\ \ea \right)$
~~& $T(n,a,b,c)=\left(\ba{ccc}
0&0&\et^a\\
0&\et^b&0\\
\et^c&0&0\\ \ea \right)$~~&
$ U(n,a,b,c)=\left(\ba{ccc}
0&\et^a&0\\
\et^b&0&0\\
0&0&\et^c\\ \ea \right)$ \\
$ R(n,a,b,c)=\left(\ba{ccc}
0&0&\et^a\\
\et^b&0&0\\
0&\et^c&0\\ \ea \right)$ 
~~&
$V(n,a,b,c)=\left(\ba{ccc}
0&\et^a&0\\
0&0&\et^b\\
\et^c&0&0\\ \ea \right)$
~~& $W(n,a,b,c)=\left(\ba{ccc}
\et^a&0&0\\
0&\et^b&0\\
0&0&\et^c\\ \ea \right)$\\
 \hline
 \hline
  \multicolumn{3}{c}{Generators of finite DSG of $U(3)$ of category (Y)}\\
 \hline
$X_1=\left(\ba{ccc}
0&\frac{1}{\sqrt{2}}\gm^{11}&\frac{1}{\sqrt{2}}\gm^{14}\\
\frac{1}{\sqrt{2}}\gm^{5}&\frac{1}{2}\gm^{20}&\frac{1}{2}\gm^{11}\\
\frac{1}{\sqrt{2}}\gm^{14}&\frac{1}{2}\gm^{17}&\frac{1}{2}\gm^{8}\\
\ea \right)$
~~& $X_2=\left(\ba{ccc}
\frac{1}{\sqrt{3}}\gm^{21}&\frac{1}{\sqrt{6}}\gm^{16}&\frac{1}{\sqrt{2}}
\gm^{13}\\
\sqrt{\frac{2}{3}}\gm^{14}&\frac{1}{2\sqrt{3}}\gm^{21}&\frac{1}{2}\gm^{
18} \\
0&\frac{\sqrt{3}}{2}\gm^{18}&\frac{1}{2}\gm^{3}\\
\ea \right)$
~~& $X_3=\left(\ba{ccc}
\frac{1}{\sqrt{3}}\vt^{31}&\frac{1}{\sqrt{6}}\vt^{14}&\frac{1}{
\sqrt{2}}
\vt^{4}\\
\sqrt{\frac{2}{3}}\vt^{30}&\frac{1}{2\sqrt{3}}\vt^{31}&\frac{1}{2}
\vt^{21} \\
0&\frac{\sqrt{3}}{2}\vt^{32}&\frac{1}{2}\vt^{4}\\
\ea \right)$ \\
$X_4=\left(\ba{ccc}
0&\frac{1}{\sqrt{2}}\vt^{13}&\frac{1}{\sqrt{2}}\vt^{12}\\
\frac{1}{\sqrt{2}}\vt^{35}&\frac{1}{2}\vt^{24}&\frac{1}{2}\vt^
{5} \\
\frac{1}{\sqrt{2}}\vt^{18}&\frac{1}{2}\vt^{25}&\frac{1}{2}\vt^
{6} \\
\ea \right)$
~~& $X_5=\left(\ba{ccc}
\frac{1}{\sqrt{3}}\ph^{9}&\sqrt{\frac{2}{3}}&0\\
\sqrt{\frac{2}{3}}\ph^{2}&\frac{1}{\sqrt{3}}\ph&0\\
0&0&\ph^5\\
\ea \right)$
~~& $X_6=\left(\ba{ccc}
\gm^{22}&0&0\\
0&\frac{1}{2}\gm^{10}&\frac{\sqrt{3}}{2}\gm^{11}\\
0&\frac{\sqrt{3}}{2}\gm^{21}&\frac{1}{2}\gm^{10}\\
\ea \right)$\\
$
X_7=\left(\ba{ccc}
\frac{1}{\sqrt{3}}\ps^{9}&\frac{1}{\sqrt{6}}\ps^{2}&\frac{1} {\sqrt{2}}
 \ps^{7}\\
 \frac{1}{\sqrt{6}}\ps^{4}&\frac{9+\sqrt{3}i}{12}&\frac{1}{2}\ps^{10}\\
 \frac{1}{\sqrt{2}}\ps^{11}&\frac{1}{2}&\frac{1}{2}\ps^{2}\\
 \ea \right)$
 ~~& $X_8=\left(\ba{ccc}
 \frac{1}{\sqrt{3}}\ps^{6}&\sqrt{\frac{2}{3}}\ps&0\\
 \sqrt{\frac{2}{3}}\ps^{11}&\frac{1}{\sqrt{3}}&0\\
 0&0&\ps^3\\
 \ea \right)$
 ~~& $X_9=\left(\ba{ccc}
 \frac{1}{\sqrt{3}}\gm^{13}&\sqrt{\frac{2}{3}}\gm^{14}&0\\
 \sqrt{\frac{2}{3}}\gm^{12}&\frac{1}{\sqrt{3}}\gm&0\\
 0&0&\gm^{19}\\\ea \right)$ \\
 $X_{10}=\left(\ba{ccc}
0&\frac{1}{\sqrt{2}}\gm^{3}&\frac{1}{\sqrt{2}}\gm^{19}\\
\frac{1}{\sqrt{2}}\gm&\frac{1}{2}\gm^{2}&\frac{1}{2}\gm^{6}\\
\frac{1}{\sqrt{2}}\gm^{21}&\frac{1}{2}\gm^{10}&\frac{1}{2}\gm^{14}\\
\ea \right)$
~~& 
~~& \\
\hline\hline
 \end{tabular}
 \end{math}
 \end{footnotesize}
\end{center}
\caption{Generators of finite DSG of $U(3)$ of order $<512$ as listed in
\cite{Ludl:2010bj}. Here $a,b,c=0,1,...,n-1$, $\et=e^{2 \pi i/n}$, $\gm=e^{2 \pi i/24}$,
$\vt=e^{2 \pi i/36} $, $\ph=e^{2 \pi i/16}$ and $\ps=e^{2 \pi i/12}$.}
\label{generators}
\end{table}

All the groups which can be generated using set (X) are not known but  the
groups with order $<512$ generated by the set (X) and (Y) and which cannot be written as direct
product of $F\times Z_n$ are listed by Ludl \cite{Ludl:2010bj}. One can obtain groups $F\times Z_n$
from this set as discussed by Ludl. There are 75 groups with order $<512$ of which only 5 fall in
the category (Y). In addition to these 75 groups,  Ludl has also identified \cite{Ludl:2010bj}
infinite series of groups called $S_4(n)$, $T_m(n)$, $\Delta(n,j)$, $\Delta(n,j,k)$ which are
expressible as semi-direct product of DSG of $SU(3)$ with some cyclic group $Z_m$. All these series
fall in the category (X).

\subsection{Massless state: General flavour structure and implications} 
\label{analytical-massless}
We now state and derive a general theorem regarding the structure of the massless state that holds
for a large class of DSG of $U(3)$. \\ 
\noindent {\bf Theorem:} If $S_{1\nu}$ and $T_l$ are elements of any group $G_f$ in category (X)
generated by combination of matrices $R,~S,~T,~U,~V,~W$ listed in Table \ref{generators} and if all
the eigenvalues of $T_l$ are distinct then the only possible flavour compositions of the massless
state are (1) trivial $(1,0,0)$ and its permutation or (2) trimaximal with probability 
$\frac{1}{3}(1,1,1)$ or (3) BM with probability $\frac{1}{2}(0,1,1)$ or permutations
thereof.\\ 
\noindent
{\bf Proof:} The proof of this theorem follows from few simple observations. We list these and
their implications in the following.
\begin{enumerate} [(1)]
 \item Let us define  generalized textures with elements in set ${\cal H}=\{R,S,T,U,V,W\}$ replaced
by ${\cal \tilde{H}}=\{\tilde{R},....,\tilde{W}\}$ obtained by replacing non-zero entries in
elements of ${\cal H}$ by some arbitrary roots of unity. Thus, for example 
\be \label{tilder}
\tilde{R}\equiv\left(\ba{ccc}
0&0&\eta_1\\
\eta_2&0&0\\
0&\eta_3&0\\ \ea \right) \ee
with $|\eta_i|=1$. ${\cal H}$ is clearly contained in  ${\cal \tilde{H}}$. It is easy to show that
product of any two elements in ${\cal H}$ is an element in ${\cal \tilde{H}}$, e.g.
$$R(n_1,a,b,c)~~ R(n_2,p,q,r)=\left(
\ba{ccc}
0&\eta_1^a\eta_2^r &0\\
0&0&\eta_1^b\eta_2^p\\
\eta_1^c\eta_2^q&0&0\\
\ea\right) \sim \tilde{V} \in {\cal \tilde{H}}~.$$
Also product of two elements in ${\cal \tilde{H}}$ belongs to ${\cal \tilde{H}}$. This implies that
{\it all}  the elements of {\it any } group $G_f$ generated using elements in ${\cal H}$
possesses only one of the six textures given in ${\cal \tilde{H}}$. It is useful to define
subsets ${\cal \tilde{H}}_1=\{\tilde{W}\}$, ${\cal \tilde{H}}_2=\{\tilde{R},\tilde{V}\}$ and
${\cal \tilde{H}}_3=\{\tilde{S},\tilde{T},\tilde{U}\}$. A product of two elements of ${\cal
\tilde{H}}_1$ belongs to ${\cal \tilde{H}}_1$ only while a product of any two elements of
${\cal \tilde{H}}_2$ generates an element which is either in ${\cal \tilde{H}}_1$ or in
${\cal \tilde{H}}_2$. A product of any two elements in ${\cal \tilde{H}}_3$ can belong to either of
${\cal \tilde{H}}_{1,2,3}$\footnote{The set of above textures is isomorphic to the $S_3$ group with 
$\tilde{W},\tilde{S},\tilde{T},\tilde{U},\tilde{R},\tilde{V}$ mapped  respectively to elements 
$e,(23),(13),(12),(132 ),(123)$ of $S_3$ \cite{Ishimori:2010au}.}.

\item Since $\tilde{W}$ is diagonal it can either be used as $S_{1\nu}$ (if  one diagonal element is
complex and two are unequal and real) or as $T_l$ (if all diagonal elements are different). Thus
\emph{${\cal \tilde{H}}_1$ can either be used as a symmetry of the neutrinos or that of the charged
leptons.}

\item Eigenvalues of $\tilde{A}=\tilde{R},\tilde{V}$ are given by $$
({\rm Det.}\tilde{A})^{\frac{1}{3}} (1,\omega,\omega^2)~ $$ with $\omega=e^{2 \pi i/3}$. Thus at
least two eigenvalues are complex and any element of $G_f$ with these two structures cannot be used
as neutrino symmetry $S_{1\nu}$. \emph{${\cal \tilde{H}}_2$ can only be used as symmetry of the
charged leptons.}

\item Each element in ${\cal \tilde{H}}_3$ contains a diagonal and two off-diagonal non-zero entries. Their eigenvalues are of two types. If the off-diagonal entries are complex conjugate of each
other then eigenvalues are given by $(\beta,1,-1)$. This can be used as neutrino symmetry
$S_{1\nu}$ if $\beta\not=\pm 1$. Otherwise elements in ${\cal \tilde{H}}_3$ also have at least two
complex eigenvalues and they would only be suitable as representing $T_l$. Thus \emph{${\cal
\tilde{H}}_3$ can be used as a symmetry of the neutrinos and/or charged leptons depending on its
structure.}
\end{enumerate}

It follows from (2-4) that if flavour group $G_f$ belongs to category (X)
then  there are two
possible choices for the neutrino symmetry $S_{1\nu}$ and three for the the charged lepton
symmetry $T_l$ apart from their cyclic permutations. We summarize below
these choices and  matrices which diagonalize them:
\begin{enumerate}[(S.a)]
 \item $S_{1\nu} \in {\cal \tilde{H}}_1$: $S_{1\nu}\equiv S_{1\nu}^a=\tilde{W}$ and
$V_\nu\equiv V_\nu^a=1.$
  \item $S_{1\nu} \in {\cal \tilde{H}}_3$: For example,
  $$S_{1\nu}\equiv S_{1\nu}^b=\left(\ba{ccc}
 \beta&0&0\\
 0&0&z^*\\
 0&z&0\\
  \ea \right)~~{\rm and}~~V_\nu\equiv V_{\nu}^b=\frac{1}{\sqrt{2}}\left( \ba{ccc}
 \sqrt{2}&0&0\\
 0&1&-z^*\\
 0&z&1\\
 \ea \right).$$ 
 \end{enumerate}
Other choices of $S_{1\nu}$ are obtained from the cyclic permutation of above.
\begin{enumerate}[(T.a)]
 \item $T_l \in {\cal \tilde{H}}_1$: $T_l\equiv T_{l}^a=\tilde{W}$ and $V_l\equiv V_l^a=1.$
 \item $T_l \in {\cal \tilde{H}}_2$: For example, $T_l\equiv T_{l}^b=\tilde{R}$ ~and~ $V_l\equiv
V_l^b={\rm Diag.}(1, \eta_2,\eta_1^*)
U_\omega.$
\item $T_l \in {\cal \tilde{H}}_3$: For example,
$$T_l\equiv T_l^c=\left(\ba{ccc}
\beta&0&0\\
0&0&z_1\\
0&z_2&0\\
\ea \right)~~{\rm and}~~V_l\equiv V_l^c=\frac{1}{\sqrt{2}}\left( \ba{ccc}
\sqrt{2}&0&0\\
0&1&-(z_1 z_2^*)^{\frac{1}{2}}\\
0&(z_2 z_1^*)^{\frac{1}{2}}&1\\
\ea \right)$$
\end{enumerate}
where $\tilde{R}$ is defined in Eq. (\ref{tilder}) and 
\be \label{uw}
U_\omega=\frac{1}{\sqrt{3}}\left( \ba{ccc}
1&1&1\\
1&\omega^2&\omega\\
1&\omega&\omega^2\\
\ea \right) ~,\ee
$\omega=e^{2 \pi i/3}$. Other choices of $T_{l}$ are obtained from the cyclic permutation of above.

The above alternatives for $S_{1\nu}$ and $T_l$ determine complete mixing pattern (apart from cyclic
permutations) which is possible for all the flavour groups belonging to category (X). The detailed
values of the mixing angles depend upon the values of complex parameters appearing above but
flavour content of the  massless state in all the cases is not sensitive to
them. To see this, note that eigenstate of
$S_{1\nu}$ corresponding to a massless neutrino for all the choices listed
above is given uniquely by
$(1,0,0)^T$ or its cyclic permutations and let us consider \state{$\psi_0$}$=(0,0,1)^T$ for
definiteness. The corresponding flavour state in the basis with diagonal $T_l$ (and hence diagonal
charged lepton mass matrix $M_l M_l^\dagger$) is given
by \state{$\psi$}$=V_l^\dagger$\state{$\psi_0$}. Then three possible choices for $V_l$ listed
above give three flavour mixing mentioned in the theorem stated above. (a) $V_l=V_l^a$ implies an
unmixed massless \state{$\psi$} with probabilities $(0,0,1)$ or (b) $V_l=V_{l}^b$ implies
trimaximally mixed \state{$\psi$}  with probabilities $\frac{1}{3}(1,1,1)$ or (c) $V_l=V_{l}^c$
implies bimaximally mixed \state{$\psi$} with probabilities
$\frac{1}{2}(0,1,1)$. Other choices allowed by permutations of 
the chosen  $S_{1\nu}$ and $T_l$ only permute the above mentioned flavour
compositions.

The above result has strong phenomenological implications due to the fact that the structure of the
massless state determines  one column of the mixing matrix. If only one neutrino is massless then
the massless state has to be identified with the first (third) column of the neutrino mixing matrix
in the standard convention in case of the normal (inverted) hierarchy. Then, for the trimaximal
structure, one predicts $\sin^2\theta_{13}=\frac{1}{3}$ for the inverted and
$\cos^2\theta_{13}\cos^2\theta_{12}=\frac{1}{3}$ for the normal hierarchy. Both of these differ from
the experimental values and would need large corrections. The BM composition is more suitable
to describe inverted hierarchy and taken as the third column of $\Upmns$ would predict
$\theta_{23}=\frac{\pi}{4}$ and $\theta_{13}=0$. These predictions signaling \mt symmetry have often
been considered as a good zeroth order ansatz which can follow in a large number of groups in
category (X) as we will show. But  with either choice, one does not get the correct leading order
value for the solar mixing angle in the whole class of groups in the category (X).

\subsection{Mixing patterns with a massless neutrino}
The foregoing discussion also allows us to determine all possible mixing matrices for all the groups
in category (X). We list below non-trivial choices.

\begin{enumerate}[(1)]
 \item If $S_{1\nu}\in{\cal \tilde{H}}_1$ and $T_l\in {\cal \tilde{H}}_2$ then $\Upmns$ has
democratic structure:
\be \label{democratic}
|\Upmns|=\frac{1}{3} \left(\ba{ccc}
1&1&1\\ 1&1&1\\ 1&1&1\\ \ea \right) \ee
The above structure which can be realized for both normal and
inverted hierarchy  implies
$\sin^2\theta_{13}=\frac{1}{3},\sin^2\theta_{12}=\sin^2\theta_{23}=\frac{1}{2}$. It fails to predict
two of the mixing angles correctly at the leading order and would need very large corrections. 

\item If $S_{1\nu}$ is to be non-diagonal then it is given by the choice (S.b). With $T_l=T_l^b$ as
in (T.b), one gets $\Upmns$ as
\be\label{case2} 
 \Upmns= \frac{1}{\sqrt{6}}
\left( \ba{ccc}
\sqrt{2}& p\eta_2^*+p^*\eta_1 z^*&z(p^* \eta_1 z^*-p \eta_2^*)\\
\sqrt{2}&p\eta_2^* \omega +p^*\eta_1 z^*\omega^2&z(p^*\eta_1
z^*\omega^2-p\eta_2^* \omega)\\
\sqrt{2}&p\eta_2^* \omega^2 +p^*\eta_1 z^*\omega&z(p^*\eta_1
z^*\omega-p\eta_2^* \omega^2)\\
\ea \right) \ee
where, $p=(\eta_1\eta_2\eta_3)^{\frac{1}{3}}$. Note that the first column determining massless state
is fixed for the normal hierarchy while one can permute the second and third columns by
interchanging the order of the two real eigenvalues of $S_{1\nu}$. Similarly, changing the order of
the three rows amount to reordering the eigenvalues of $T_l$. Thus there are six possible choices
for the reactor mixing angle $\theta_{13}$ and minimum of this should be identified with
$\sin^2\theta_{13}$ and the corresponding column as the third column of the PMNS matrix. Explicitly,
\be \label{s13}
\sin^2\theta_{13}= \frac{1}{3}~{\rm Min.~}|1\pm {\rm Re}(\eta_3^*p^*\tilde{\lambda_i}z)|~,\ee
where $\pm$ sign correspond to two columns and rows are labeled by $$\tilde{\lambda}_i=
(1,\omega,\omega^2)~.$$ 
$\eta_3$ is defined in Eq. (\ref{tilder}). The $\sin^2\theta_{23}$ also gets
determined by the entries in the same column as that of $\sin^2\theta_{13}$. Moreover, it is seen
from Eq. (\ref{case2}) that
$$ {\rm Im}(U_{12}U_{13}^*U_{23}U_{22}^*)=0$$
independent of the values of complex parameters showing that all the groups in category (X) lead to
the absence of the Dirac CP violation even if they predict non-zero $\theta_{13}$. Alternative
choice $T_l=\tilde{V}$ also give prediction similar to above but other possible choices of $T_l$, 
namely $T_l \in {\cal \tilde{H}}_3$ give unacceptable predictions for mixing angles in case of the
normal hierarchy.
\item Unique choice for $S_{1\nu}$ in case of the inverted hierarchy is given by
\be S_{1\nu}=\tilde{U}=
\left( \ba{ccc}
0&z_\nu&0\\
z_\nu^*&0&0\\
0&0&\beta_\nu\\
\ea \right)~.\ee
The corresponding choice for $T_l$ giving non-trivial pattern is given by $T_l^c$ in (T.c).
The resulting mixing matrix is given by:

\be \Upmns=
\left( \ba{ccc}
\frac{1}{\sqrt{2}}&-\frac{z_{\nu}}{\sqrt{2}}&0\\
\frac{z_{\nu}^*}{2}&\frac{1}{2}&\frac{(z_{2}z_{1}^*)^{\frac{1}{2}}}{\sqrt
{2}}\\
-\frac{z_{\nu}^*(z_{1}z_{2}^*)^{\frac{1}{2}}}{2}&-\frac{(z_{1}z_{2}^*)^
{ \frac{1}{2}}}{2}&\frac{1}{\sqrt{2}}\\
\ea \right) \ee
Note that the $|\Upmns|$ has the BM values $\theta_{12}=\theta_{23}=\frac{\pi}{4}$ and
$\theta_{13}=0$ independent of the values of the group dependent complex parameters. In this sense,
the BM mixing pattern is more universal and follows in many groups constructed from
the category (X).
\end{enumerate}

The case (2) discussed above can lead to reasonable values for the mixing angles $\theta_{13}$,
$\theta_{23}$ and we will explore various possibilities numerically in the next section. The
$\theta_{12}$ is predicted either $\theta_{12}=\frac{\pi}{4}$ or close to $\theta_{12}\sim
54^{\circ}$. This needs to be changed and we shall discuss possible ways to modify this in section
\ref{correction}.

Before closing this section, we wish to emphasize generality of the result derived here. Firstly,
this result is valid for any group $G_f$ in category (X)  independent of its
order. Secondly, large number of the finite DSG of $U(3)$
fall in category (X)
to which the theorem derived here is applicable. In addition to 70 of the 75 groups with order
$<512$, six infinite series of groups also fall in this category. These are generated by the
following specific matrices of Table \ref{generators}:
\beqa 
\Sigma(3 n^3)&\equiv&\{W(n,0,0,1),~W(n,1,0,0),~E\}~; \nonumber \\
T_N(m)&\equiv&\{e^{2\pi i/3^m} E,~W(n,1,a,n-1-a)\}~,~~~(1+a+a^2)~{\rm mod}~n=0~; \nonumber \\
S_4(m)&\equiv&\{S(2^m,1,1+2^{m-1},1),~T(2^m,3,3,2^{m-1}+3)\}~; \nonumber \\
\Delta(3n^2,m)&\equiv&\{e^{2\pi i/3^m} E,~W(n,0,1,n-1)\}~; \nonumber \\
\Delta(6 n^2,m)&\equiv&\{E,~W(n,0,1,n-1),~T(2^m,1+2^{m-1},1+2^{m-1},1+2^{m-1})\}~; \nonumber \\
\Delta(6 n^2,j,k)&\equiv&\{E,~W(n,0,1,n-1),~T(3^j 2^k,1+3^j2^{k-1},1+3^j2^{k-1},1+3^j2^{k-1})\}.
\nonumber \eeqa
$E$ defined above is a matrix with determinant 1 and can be written as $E=V(p,0,0,0)$ independent of
$p$. It is clear from the structure of the generators that the  group series
$\Sigma(3n^3),~T_n(m),~\Delta(3 n^2,m)$ either contain diagonal elements or elements with two
complex eigenvalues. Thus these groups provide an example of the case (1) discussed before Eq.
(\ref{democratic}) and only mixing pattern possible within these groups is
democratic mixing. Of these, $\Sigma(3 n^3)$ were already discussed in
\cite{Joshipura:2013pga}.

The above theorem does not however hold for the groups in category (Y). These are not easily
amenable to analytic discussion but we study them numerically and show that
one of these can give
quite satisfactory leading order predictions for the mixing angles.

\section{A numerical scan of discrete subgroups of $U(3)$}
\label{scan}
We now study numerically the predicted values of the reactor and atmospheric mixing angles for
all the DSG of $U(3)$ of order $<512$. These include all the 70 groups of category (X) and 5 groups
of category (Y). It is clear form the discussions in the last section that if the groups of category
(X) are used as flavour symmetry for a massless neutrino then sizable corrections at least in the
solar angle would be needed which in general may change the other angles also. It is thus
appropriate to work out leading order predictions for other angles. While the predicted solar angle
is almost universal for all the groups under study, the other angles depend on the details of the
group and we present this numerically. In our analysis, we choose a particular group and look at its
suitability to be a flavour symmetry with a massless neutrino and work out all possible values of
the predicted mixing angles.

\subsection{Results of the groups in category (X)}
First, we restrict ourselves to the groups of category (X) and of order $<512$  tabulated by Ludl
\cite{Ludl:2010bj}. We demand that (1) this groups should contain at least one element to be
identified as $S_{1\nu}$ with eigenvalue $(\eta,1,-1)$ with $\eta^N=1$ and $N\geq 3$ appropriate to
describe a massless neutrino, (2) $T_l$ should have all three different eigenvalues and (3)
$S_{1\nu}$ does not commute with $T_l$. These requirements are essential but quite restrictive and
rule out large number of groups as possible $G_f$. Consider all the groups generated only from the
matrices $R$, $V$ and $W$ alone for arbitrary values of the integers defining these matrices. As
mentioned in the last section, the above structures close among themselves as result the groups in
question contain either diagonal elements or elements with at least two complex eigenvalues. The
neutrino symmetry in this case would come only from the diagonal elements and one would get only
democratic mixing which is not of much interest.  There are 48 such groups among the 70 listed
groups and the same argument would apply to any higher order groups generated using any of these
three matrices. Requirement (1) is also restrictive and we find numerically that  of the remaining
22 groups, 16 get ruled out by this requirement. This leaves only six groups which can be used as
suitable flavour symmetry. We numerically  find all the elements which can be used as $S_{1\nu}$ and
$T_l$ within these allowed groups and determine the resulting mixing matrix. The allowed groups
and predicted values of the mixing angles are listed in Table \ref{NHresults}. 

\begin{table}[!ht]
\begin{small}
\begin{center}
\begin{tabular}{cccccc}
 \hline
 \hline
  Group & Classification & Generators & $|\Upmns|$ ~~&~~ $\sin^2\theta_{13}$ ~~&~~
  $\sin^2\theta_{23}$\\
 \hline
 \group{48,30}  &$S_4(2)$ & $S(4,1,3,1),~T(4,3,3,1)$ & $M_1$ &0.0447 &0.349\\
 \group{162,10} &  & $S(3,0,1,0),~T(3,1,1,0)$ &$M_2$& 0.0201 & 0.399\\
                &  &                          &$M_3$& 0 & 0.5\\
 \group{192,182} & $\Delta(6\cdot4^2,2)$ & $T(4,0,2,1),~U(4,3,0,2)$&$M_1$&0.0447 & 0.349\\
                &   &   &$M_3$& 0 & 0.5\\
 \group{384,571} & $\Delta(6\cdot4^2,3)$ & $T(8,1,5,3),~U(8,1,3,7)$&$M_4$&0.0114 & 0.424\\
 \group{432,260} &$\Delta(6\cdot6^2,2)$ & $S(12,7,5,3),~T(12,9,5,7),~U(12,9,1,11)$ &$M_1$&0.0447 &
            0.349\\
 \group{486,125} & & $S(9,2,5,2),~T(9,7,7,4),~U(3,0,1,1)$ &$M_2$& 0.0201 & 0.399\\
                & &   &$M_3$& 0 & 0.5\\
\hline
\hline
\end{tabular}
\end{center}
\end{small}
\caption{Predictions for $\theta_{23}$ and $\theta_{13}$ for the normal hierarchy ($m_{\nu_1}=0$)
from a scan of the DSG of $U(3)$ of category (X) and order $<512$. The group identity \group{$g, j$}
denotes the $j^{\rm th}$ finite group of order $g$ as classified in the Small Group Library of the
computer algebra system GAP \cite{GAP4}. The prediction for $\sin^2\theta_{23}$ is chosen to be in
the first octant of $\theta_{23}$. The structures $M_1$-$M_4$  for $\Upmns$
are specified in Eq. (\ref{UPMNS-structure}).}
\label{NHresults}
\end{table}

We have restricted ourselves only to choices which give $\sin^2\theta_{13}\leq (0.25)^2$. In this
case one gets only four possible forms of $|\Upmns|$ in case of the normal neutrino mass hierarchy.
They are  listed as $M_1,~M_2,~M_3,~M_4$ below:
\be \label{UPMNS-structure}
\ba{cc}
M_1=\left(
\begin{array}{ccc}
 0.5774 & 0.7887 & 0.2113 \\
 0.5774 & 0.5774 & 0.5774 \\
 0.5774 & 0.2113 & 0.7887 \\
\end{array}
\right)~;
&~~~
M_2=\left(
\begin{array}{ccc}
 0.5774 & 0.8041 & 0.1418 \\
 0.5774 & 0.5248 & 0.6255 \\
 0.5774 & 0.2793 & 0.7673 \\
\end{array}
\right)~; \nonumber \\
\ea
\ee
\be
\ba{cc}
M_3=\left(\ba{ccc}
\sqrt{\frac{1}{3}}&\sqrt{\frac{2}{3}}&0\\
\sqrt{\frac{1}{3}}&\sqrt{\frac{1}{6}}&\sqrt{\frac{1}{2}}\\
\sqrt{\frac{1}{3}}&\sqrt{\frac{1}{6}}&\sqrt{\frac{1}{2}}\\
\ea
\right)~;&~~~~
M_4=\left(
\begin{array}{ccc}
 0.5774 & 0.8095 & 0.1066 \\
 0.5774 & 0.4971 & 0.6478 \\
 0.5774 & 0.3125 & 0.7543 \\
\end{array}
\right)~.\\
\ea
\ee

Note that there exist two predictions for $\theta_{23}$ in each of the above pattern as the second
and third row can be interchanged though in Table \ref{NHresults} we select its first octant
($\theta_{23} \le \pi/4$) values as they are favored at $2\sigma$ in case of normal hierarchy by a
recent global fit \cite{Capozzi:2013csa}. It is seen from the above equations that

\begin{itemize}
\item Structure $M_3$ displays \mt symmetry.
\item $M_4$ predicts $\sin^2\theta_{23}$ close to the best fit value as per the global fits in 
\cite{Capozzi:2013csa}. The predicted $\sin^2\theta_{13}$ requires significant corrections.
\item $M_2$ fares better and predicts $\sin^2\theta_{13}$ within $2\sigma$ and $\sin^2\theta_{23}$
within $1\sigma$ of their  best fit values.
\end{itemize}

All the structures have trimaximally mixed massless state and predict $\sin^2\theta_{12}\sim 
\frac{2}{3}$. Situation changes if one were to impose lepton number as an additional symmetry and
assume that neutrinos are Dirac particles. All the residual symmetries considered here and many more
will be allowed choices in this case but now none of the neutrinos would be forced to be massless.
In this case, one has the freedom to interchange the first and the second columns of all the
structures obtained here. The group \group{162,10} in this case will predict all the mixing angles
correctly within 3$\sigma$. This group has already been identified as a possible group for the Dirac
neutrinos \cite{Holthausen:2013vba}.

The above results are obtained for the normal hierarchy. The inverted hierarchy allows only  the
BM pattern as argued analytically in the previous section. All the groups listed in Table
\ref{NHresults} predicts BM mixing for the inverted hierarchy. The group $S_4(2)$ is the smallest
among them and has been studied in detail in \cite{Joshipura:2013pga} where an explicit model giving
satisfactory mixing pattern is discussed.

\subsection{Results of the groups in category (Y)}
We now discuss the groups in category (Y) with order $< 512$ listed by Ludl \cite{Ludl:2010bj}.
There are only five such groups having 3-dimensional faithful IR. The corresponding generators are
given by matrices $X_1$-$X_{10}$ and reproduced here in Table \ref{generators}.  Consecutive pairs
of matrices $X_i,~X_{i+1}$ with $i=1,3,5,7,9$ generate groups of order 216, 324, 432, 432, 432
respectively. We have numerically generated all the elements of these five groups and find that four
of the five groups do not contain any element with one complex and two real eigenvalues. They
therefore do not qualify as possible flavour groups for a massless neutrino in this approach. The
group generated by $X_7,~X_8$ (classified in GAP \cite{GAP4} as \group{432,239}) is the only one
which contain such elements and we analyze it further.

The group \group{432,239} generated numerically from the products of $X_7,~X_8$ contain 372
elements with all eigenvalues unequal. Of these, 72 have two real eigenvalues and would be
appropriate as a member of $G_\nu$. The corresponding charged lepton symmetry group $G_l$ may be
generated from any of the 372 elements. Taking only the non-commutative set of $G_l$ and $G_\nu$,
all the mixing angles are fully predicted with any of these choices due to their completely
different eigenvalues. We have numerically constructed all possible mixing matrices using elements
in these sets. We find that this group allows only two possible structures for the massless state.
These have flavour content $\sim(0.59,0.25,0.16)$ or $(0,1/2/,1/2)$ and permutations of these. Since
the former now differs from the trimaximal structure predicted in class (X) groups, there is a
possibility of better agreement with the data. We explore it further by demanding that the reactor
mixing angle should satisfy $|\sin\theta_{13}|\leq 0.25$. It turns out that only two possible mixing
structures satisfy this restriction, one for the normal and one for the inverted hierarchy. These
are respectively given by,
\be \label{normalux6x7}
|\Upmns|=\left(
\begin{array}{ccc}
 0.7691 & 0.6124 & 0.183 \\
 0.3981 & 0.683 & 0.6124 \\
 0.5 & 0.3981 & 0.7691 \\
\end{array}
\right)~~{\text{for normal hierarchy}}\ee
and
\be \label{invertedux6x7}
|\Upmns|=\left(
\begin{array}{ccc}
 0.8881 & 0.4597 & 0 \\
 0.3251 & 0.628 & 0.7071 \\
 0.3251 & 0.628 & 0.7071 \\
\end{array}
\right)~~{\text{for inverted hierarchy}}.\ee

The predicted mixing angles are: 
\beqa \label{prediction432}
\sin^2\theta_{12}&=&0.388,~\sin^2\theta_{23}=0.388,~\sin^2\theta_{13}=0.0335~{\rm
for~normal~hierarchy}, \nonumber\\
\sin^2\theta_{12}&=&0.211,~\sin^2\theta_{23}=1/2,~\sin^2\theta_{13}=0~{\rm for~inverted~~hierarchy}
~.\eeqa
The prediction in case of the normal hierarchy does not reproduce all the mixing angles within
3$\sigma$. But it is fairly close to the observed ones with $\chi^2\sim 45$. It is thus conceivable
that small non-leading order effect can correct the predictions. The same predictions can be
obtained for all massive neutrinos if they are of Dirac type\footnote{The same pattern is
obtained for Dirac neutrinos in \cite{Hagedorn:2013nra} using the group $\Sigma(36 \times 3)$ which
is an $SU(3)$ subgroup of \group{432,239} discussed here.}. The predictions in case of the inverted
hierarchy are similar in some sense to the BM mixing. The predicted $\theta_{23}$ and $\theta_{13}$
are the same as in the BM mixing. The solar angle predicted here requires large corrections which
are similar in magnitude to the case of the BM mixing but now in opposite direction.

The above mixing matrices can arise with several different choices of $S_{1\nu}$ and $T_l$ from 
elements of the group. We give here an example for the normal hierarchy which happens to be the
simplest choice:
\be
T_l=X_7 X_8=
\left(
\begin{array}{ccc}
 \frac{i}{3}+\frac{1}{3} e^{\frac{i \pi }{6}} & -\frac{1}{3} i \sqrt{2} e^{\frac{i \pi }{6}}+
\frac{e^{\frac{i \pi }{3}}}{3 \sqrt{2}} &
   \frac{i e^{-\frac{5 i \pi }{6}}}{\sqrt{2}} \\
 \frac{\left(9+i \sqrt{3}\right) e^{-\frac{i \pi }{6}}}{6 \sqrt{6}}- \frac{e^{\frac{2 i \pi }{3}}}{3
\sqrt{2}} & \frac{9+i \sqrt{3}}{12\sqrt{3}}+\frac{1}{3} e^{\frac{5 i \pi }{6}} & \frac{1}{2} i
e^{-\frac{i \pi }{3}} \\ 0 & \frac{\sqrt{3}}{2} & \frac{1}{2} i e^{\frac{i \pi }{3}} \\
\end{array}
\right)\ee

\be
S_{1\nu}=X_8 X_7=
\left(
\begin{array}{ccc}
 \frac{i}{3}+\frac{1}{3} e^{\frac{5 i \pi }{6}} & \frac{\left(9+i \sqrt{3}\right) e^{\frac{i \pi }
 {6}}}{6 \sqrt{6}}-\frac{e^{\frac{i \pi }{3}}}{3 \sqrt{2}} & \frac{e^{-\frac{i \pi
}{6}}}{\sqrt{6}}-\frac{e^{-\frac{5 i \pi }{6}}}{\sqrt{6}} \\
 -\frac{1}{3} i \sqrt{2} e^{-\frac{i \pi }{6}}+\frac{e^{\frac{2 i \pi }{3}}}{3 \sqrt{2}} & 
 \frac{9+i \sqrt{3}}{12 \sqrt{3}}+\frac{1}{3}
   e^{\frac{i \pi }{6}} & -\frac{1}{\sqrt{3}}+\frac{e^{-\frac{i \pi }{3}}}{2 \sqrt{3}} \\
 \frac{i e^{-\frac{i \pi }{6}}}{\sqrt{2}} & \frac{i}{2} & \frac{1}{2} i e^{\frac{i \pi }{3}} \\
\end{array}
\right)
\ee
Both $S_{1\nu}$ and $T_l$ have identical set of eigenvalues $(i,1,-1)$ and each therefore generate
a $Z_4$ subgroups of the full group. Diagonalization of $S_{1\nu}$ and $T_l$ generates the mixing
matrix given in Eq. (\ref{normalux6x7}). One also finds trivial CP phase and therefore absence of CP
violation in neutrino oscillations.

\section{Modifying Solar Mixing angle: Two alternatives}
\label{correction}
Since most of the groups except the group \group{432,239} lead to unsatisfactory predictions for the
solar mixing angle, it is appropriate to explore  possibilities which allows one to replace or avoid
such prediction. We propose  two alternatives, (1) use of discrete groups admitting reducible (2+1)-dimensional faithful representation such that one of the three mixing angles remains unpredicted and (2) modifications in the neutrino mass  matrices predicted at the leading order in groups in class (X).

\subsection{Discrete subgroups of $U(2)\times U(1)$}
We now depart from the requirement of assigning leptons to 3-dimensional IR and consider instead
alternative possibility of putting them into a reducible (2+1)-dimensional faithful representation. Such groups will be subgroup of
$U(2)\times U(1)$ \cite{Ludl:2010bj, Parattu:2010cy}. As we will see, consideration of such groups
allows elements having the structure of $S_{1\nu}^\prime$, Eq. (\ref{zn}) with two degenerate
eigenvalues. One mixing angle corresponding to rotation in the plane of the degenerate eigenvalues
remain undetermined in this case. Thus this scheme is less predictive but can avoid wrong prediction
of the solar mixing angle.

Let us consider again the set of groups generated from the matrices in Table \ref{generators}. One
would generate subgroups of $U(2)\times U(1)$ from this set if they are generated from
block-diagonal matrices only. Thus combination of two different $S$ or $S$ and $W$
would\footnote{note that product of two $S$ generate a $W$ like structure or product of $S$, $W$
generate a structure similar to $S$.} generate such groups and the same would apply when
$S$ is replaced by $T$ or $U$. There exist three  different possibilities for choosing neutrino and
charged lepton symmetries from the elements of the group generated this way. (a) All the eigenvalues
of the neutrino and the charged lepton symmetry generators are different, (b) two of the real
eigenvalues of the neutrino symmetry are degenerate and the charged lepton symmetry has three
distinct eigenvalues and (c) neutrino symmetry is characterized by different eigenvalues  and the
charged lepton symmetry has two degenerate eigenvalues. It is easy to see that possibility (a)
predicts two of the mixing angles to be zero. In case of possibility (b) only reasonable mixing
pattern is found with the choice
\be \label{mutau}
S_{1\nu}'={\rm Diag.}(1,1,\eta)~~~{\rm and}~~~ T_l=\left(\ba{ccc}
\eta_1&0&0\\
0&0&\eta_2\\
0&\eta_3&0\\ \ea \right)~. \ee
This case corresponds to the inverted hierarchy and the mixing pattern given by
$$\Upmns=U_{23}(\pi/4)U_{12}(\theta)~,$$
where $U_{ij}(\theta)$ is a unitary rotation in the $ij$-plane with an angle $\theta$ and 
undetermined phases. The above mixing matrix corresponds to the \mt symmetric structure and predict
$\theta_{13}=0$ and $\theta_{23}=\pi/4$. The small next to leading order effects are known to
generate large enough $\theta_{13}$ in this case \cite{Gupta:2013it}. In case (c), the atmospheric
mixing angle remains undetermined and the solar angle gets fixed to $\pi/4$.

The $S_{1\nu}'$ and $T_l$ given above close to a finite group if $\eta$, $\eta_{1,2,3}$ all are 
$n^{\rm th}$ roots of unity. $S_{1\nu}'$ is given in this case by appropriate $W$ in Table
\ref{generators} and $T_l$ by $S$. Now define three generators:
\be 
\ba{ccc}
A=\left(\ba{ccc}
\eta&0&0\\
0&1&0\\
0&0&1\\ \ea\right);&~~
B=\left(\ba{ccc}
1&0&0\\
0&0&1\\
0&1&0\\ \ea\right);&~~
C=\left(\ba{ccc}
1&0&0\\
0&\eta&0\\
0&0&1\\ \ea\right), \ea \ee
where $\eta=e^{2\pi i/n}$. The elements in the set ${\cal A}=\{S(n,a,b,c),~W(n,a',b',c')\}$
can be written in terms of $A$, $B$, $C$ as: 
$$ S(n,a,b,c)=A^aC^bBC^c~;~~~W(n,a',b',c')=A^{a'}C^{b'}(BCB)^{c'}$$
Thus $S$ and $W$ in the set ${\cal A}$ are elements of the group generated by $A$, $B$, $C$. It is
known that $B$, $C$ generate the group $\Sigma(2n^2)$ \cite{Ishimori:2010au, Joshipura:2013pga}.
Also $A$ commutes with $B$, $C$ and they all together generate  the group $\Sigma(2n^2)\times Z_n$.
It follows therefore that the residual symmetries in Eq. (\ref{mutau}) close to a subgroup of
$\Sigma(2n^2)\times Z_n$ which may be considered as appropriate flavour group corresponding to a
massless neutrino and \mt symmetry. Such groups without the factor $Z_n$ were considered as possible
symmetry for the massless neutrino in \cite{Joshipura:2013pga}. The $Z_n$ factor is not required if
$\eta_2\eta_3\not =1$ since in this case, all the eigenvalues of $T_l$ can be different even when
$\eta_1=1$. As the above argument shows, the group series $\Sigma(2n^2)$ or even their subgroups in
some cases may also be sufficient to harness appropriate residual symmetry.
Let us give some
examples:
\be \ba{ccc}
~~~{\rm Generators}~~~&~~~{\rm Order}~~~&~~~S_\nu'~~~\\
S(3,0,1,1),~W(3,0,0,1)&18&W\\
S(4,0,1,1),~W(4,0,0,1)&32&W\\
S(6,0,4,3),~W(6,0,3,1)&36&W^2\\
S(12,0,4,2),~W(12,0,4,1)&96&W^3\\
\ea \ee
The $S$ given in above examples can be considered as generating $G_l$ and powers of $W$ mentioned 
in the last column can be used  $S_{1\nu}'$. In the list above, the first two groups are
$\Sigma(2\cdot3^2)$ and $\Sigma(2\cdot4^2)$ while the last two are subgroups of
$\Sigma(2\cdot6^2)$ and $\Sigma(2\cdot12^2)$ respectively.

\subsection{Modified neutrino mass matrices}
All the results derived so far  are valid at the leading order when the flavour symmetry group
$G_f$ is suitably broken into $G_l$ and $G_\nu$ which imply a unique contribution to the charged
lepton and neutrino mass matrices. One can however  consider a more general possibility in which
either the charged lepton and/or the neutrino mass matrices contain two separate pieces, each
emerging separately after the spontaneous symmetry breaking of $G_f$. The
second piece may also be a leading order effect arising from a
set of flavons with different vacuum configurations or may  represent
perturbation to the leading order predictions arising  from the next
to leading order
operators which do not respect $G_l$ or $G_\nu$
The prediction of a massless neutrino will be maintained if $G_\nu$ remains intake and both these
corrections emerge in the charged lepton sector only. We have already discussed one such example in
the context of specific $S_4(2)$ model in
\cite{Joshipura:2013pga} with inverted hierarchy and BM mixing pattern. In this example, $S_4(2)$ is
broken down to $G_\nu=Z_4$ in the neutrino sector while it is completely broken in the charged
lepton sector leaving an accidental $Z_2$ symmetry, see \cite{Joshipura:2013pga} for more details.
It is found that such a correction modifies both the solar and the reactor mixing angles and brings
all three mixing angles in agreement with their global fit values at $2\sigma$ ranges.

Here we consider a scenario in which neutrino mass matrix consist of two pieces each invariant under
separate symmetries contained in $G_f$:
\be \label{two}
M_\nu=M_{1\nu}+M_{2\nu}~.\ee
Both $M_{i\nu},~(i=1,2)$ can arise from different sets of flavons after spontaneous breaking of
$G_f$ and each is assumed to be characterized by its own residual symmetries
\be \label{s12}
S_i^TM_{i\nu}S_i=M_{i\nu}~.\ee
The combined residual symmetry of the neutrino mass matrix need not be
either of $S_i$ but
the form of each $M_i$ is restricted by Eq. (\ref{s12}) and still a
predictive model which can correct for the wrong solar mixing angle can
emerge from
such breaking of $G_f$ as we shall illustrate through two different examples, one for the
normal and the other for the inverted hierarchy. If $S_1$ and $S_2$ commute then the leading order
prediction for the mixing angles would remain unaffected but the combined effect of two pieces would
be to change the mass ordering which can lead to  correct mixing angle predictions. Both mixing
angles and mass ordering can change when $S_1$ and $S_2$ do not commute. The following two examples
illustrate these two possibilities.

\subsubsection{The group \group{162,10}}
As discussed before, the group \group{162,10} generates the mixing  pattern $M_2$ in case of the
normal hierarchy. This predicts viable values of $\theta_{13}$ and $\theta_{23}$. A
satisfactory prediction of the solar mixing angle can be obtained if the first and second
columns of $M_2$ can be interchanged. However, this interchange leads to an inconsistent prediction
$m_{\nu_2}=0$. This can be corrected by admitting a second piece in $M_\nu$. 

An example of the residual symmetries of neutrinos and charged leptons which give rise to the
pattern $M_2$ in group \group{162,10} is given by
\be \ba{cc}
S_1=\left(\ba{ccc}
\omega&0&0\\
0&0&\omega\\
0&\omega^2&0\\
\ea \right)~~~{\rm and}~~~ 
T_l=\left(\ba{ccc}
0&\omega^2&0\\
0&0&1\\
1&0&0\\ \ea \right)~\\ \ea \ee
with $\omega=e^{2\pi i/3}$. The neutrino mass matrix invariant under $S_1$ takes the form 
\be \ba{cc}
M_{1\nu}=\left(\ba{ccc}
0&0&0\\
0&x&c\\
0&c&\omega^2x\\
\ea \right)~,\\ \ea \ee
where $x$ and $c$ are complex parameters. The above matrix is diagonalized by $V_\nu^b$ given in
(S.b) in section \ref{analytical-massless} with a replacement of $z$ by $\omega$.

One of the eigenvalues in lower $2\times2$ block of $M_{1\nu}$ vanishes if an ad hoc condition
$c=\pm \omega x$ is imposed. This leads to two massless neutrinos in $M_{1\nu}$ and $m_{\nu_3}=2|x|$. If the solar mass difference is generated by allowing a small perturbation in the
(1,1) entry in $M_{1\nu}$, one gets
\be \ba{cc}
M_\nu=\left(\ba{ccc}
\epsilon&0&0\\
0&x& \omega x\\
0& \omega x&\omega^2x\\
\ea \right)~,\\ \ea \ee
with $|\frac{\epsilon}{2x}|\approx\sqrt{\frac{\Dms}{\Dma}}$. The above matrix is diagonalized by 
$$U_\nu=V_\nu^b~P_{12},$$
where $P_{12}$ is a permutation matrix in 1-2 plane. This $U_\nu$ together with $U_l$ that
diagonalizes $T_l$ modify the pattern $M_2$ by interchanging its first and second columns. This
leads to $\sin^2\theta_{12}=0.34$ without perturbing the successful predictions of $\theta_{23}$
and $\theta_{13}$. In this example, the original massless neutrino associated with the
eigenvector $\frac{1}{\sqrt{3}}(1,1,1)^T$ becomes $m_{\nu_2}$ after the perturbation and the
massless state arise due to the assumption $c=\pm \omega x$. A slight departure from this
assumption leads to all massive neutrinos but without changing the mixing pattern as long as
$|c\pm\omega x| < |\epsilon|$. Perturbation $\epsilon$ is small compared to the dominant mass $2|x|$
but needs to be assumed larger than the second mass. The perturbation matrix in the above example,
{\it i.e.} $M_{2\nu}=M_\nu-M_{1\nu}={\rm Diag.}(\epsilon,0,0)$ satisfies $S_2^T M_{2\nu}
S_2=M_{2\nu}$ with $S_2={\rm Diag.}(1,\omega,\omega)$. The $S_2$ belongs to the group \group{162,10}
and it is a product of two $S(3,0,1,0)$ which is one of the generators of this group. Hence it is
possible that appropriate flavons can break the group \group{162,10} spontaneously in to two
different subgroups characterized by $S_1$ and $S_2$ and generates two contributions $M_{1\nu}$ and
$M_{2\nu}$ invariant under them, respectively. The full neutrino mass
matrix $M_\nu=M_{1\nu}+M_{2\nu}$ is not invariant under $S_{1,2}$.

\subsubsection{The group $S_4(2)$}
The $S_4(2)$ is an order 48 group generated from specific matrices $S(4,1,3,1)$ and $T(4,3,3,1)$
given in Table \ref{generators}. It is isomorphic to $A_4\rtimes Z_4$ and contains  $S_4$ as a
subgroup which has been discussed as possible symmetry for the TBM mixing \cite{reviews}. Let us
choose the following residual symmetries which are contained in $S_4(2)$:
\be \ba{ccc}
S_1=\left(\ba{ccc}
i&0&0\\
0&0&-i\\
0&i&0\\
\ea \right)~,&~~
S_2=\left(\ba{ccc}
0&0&-i\\
0&-i&0\\
i&0&0\\
\ea \right)~,&~~
T_l=\left(\ba{ccc}
0&1&0\\
0&0&1\\
1&0&0\\ \ea \right)~.\\ \ea \ee
The pieces $M_{i\nu}$ of the neutrino mass matrix  invariant under $S_{i}$
are given by:
\be \label{m1m2}\ba{cc}
M_{1\nu}=\left(\ba{ccc}
0&0&0\\
0&x_1&a_1\\
0&a_1&-x_1\\
\ea \right)~,&~~
M_{2\nu}=\left(\ba{ccc}
x_2&0&a_2\\
0&0&0\\
a_2&0&-x_2\\
\ea \right)~.\\ ~\ea \ee
The above structures of $M_{i\nu}$ may arise in specific models like the one recently proposed by us
in \cite{Joshipura:2013pga}. Assuming that the three generations of lepton doublets $L_L$ transform
as a triplet $3_3$ and a set of flavon fields $\phi_i^\nu$, $\chi_i^\nu$ ($i=1,2$) transform as
$3_2$ and $2_1$ respectively under $S_4(2)$, one can write a leading order operator for the light
Majorana neutrino masses as\footnote{We refer reader to \cite{Joshipura:2013pga, Ludl:2010bj} for
the details of $S_4(2)$ group, its representations and the relevant tensor product decomposition
rules.}
\be \label{s42-model}
{\cal W}_\nu = \sum_{i=1,2}(x'_i L_L L_L \chi_i^\nu + a'_i L_L L_L \phi_i^\nu)\frac{H_u
H_u}{\Lambda^2}~, \ee
where $H_u$ is an $S_4(2)$ singlet Higgs doublet in the minimal supersymmetric standard model. The
vacuum configuration $\langle \phi_1^\nu \rangle = \upsilon_{\phi_1}(1,0,0)^T$, $\langle \phi_2^\nu
\rangle = \upsilon_{\phi_2}(0,1,0)^T$, $\langle \chi_1^\nu \rangle = \upsilon_{\chi_1}(1,0)^T
$ and $\langle \chi_2^\nu \rangle = \upsilon_{\chi_2}(1,\sqrt{3})^T$ then leads to the forms
of $M_{i\nu}$ given in Eq. (\ref{m1m2}) with $x_i=x'_i \upsilon_{\chi_i} \langle H_u
\rangle^2/\Lambda^2$ and $a_i=a'_i \upsilon_{\phi_i} \langle H_u \rangle^2/\Lambda^2$. Similarly,
$M_lM_l^\dagger$ with the above residual symmetry can be obtained from the charged leptons Yukawa
interactions similar to the one given in \cite{Joshipura:2013pga} but with a different vacuum
structure $\langle \phi_1^l \rangle =\upsilon_{\phi_1^l}(1,1,1)^T$, $\langle \phi_2^l
\rangle=\upsilon_{\phi_2^l}(1,\omega,\omega^2)^T$
and $\langle \phi_3^l\rangle=\upsilon_{\phi_3^l}(1,\omega^2,\omega)^T$.  Note that for a given $i=1$
or 2, the vacuums of $\phi_i^\nu$ and $\chi_i^\nu$ leave the symmetry in $M_{i\nu}$ corresponding to
$S_{i\nu}$ unbroken and each piece satisfies Eq. (\ref{s12}). The combined matrix $M_\nu$ is not
invariant under either but it still leads to interesting
predictions. This follows from the fact that the neutrino mass matrix  is controlled by four
(instead of six) parameters $(x_{1,2},~a_{1,2})$ and the charged lepton mixing matrix $U_l$ given by
Eq. (\ref{uw}) does not contain any free parameters. We shall assume $x_{1,2}$, $a_{1,2}$ to be
real. In this case, the predictions of $M_\nu$ become transparent in the flavour basis with
diagonal $M_lM_l^\dagger$:
\be
M_{\nu f}\equiv U_l^TM_\nu U_l=U_\omega^TM_\nu U_\omega=\left(
\ba{ccc}
2 x&a&a^*\\
a&y&-x\\
a^*&-x&y^*\\
\ea \right)~,\ee
where $x,y,a$ are given in terms of $x_{1,2},a_{1,2}$. $x$ is real and is related to four
parameters in $y,a$ by:
\be \label{relation}
x=\frac{1}{3}({\rm Re}(y-a)+\sqrt{3} ~{\rm Im}(y+a))~.\ee
$M_{\nu f}$ is invariant under the generalized $\mu$-$\tau$ symmetry \cite{grimus} which combines CP
and the $\mu$-$\tau$ symmetry. This symmetry  appears here as an accidental symmetry of
$M_{\nu f}$ when $x_{1,2},~a_{1,2}$ are real. It is known \cite{grimus} that this symmetry leads to
equality $|U_{\mu a}|=|U_{\tau a}|$ for $a=1,2,3$ and results in two predictions:
\be \label{predictions}
\sin\theta_{13}\cos\delta=0~,~~\theta_{23}=\frac{\pi}{4}\ee
Other interesting prediction also emerges in the limiting case of real $a,y$. In this limit, $M_{\nu
f}$ obeys $\mu$-$\tau$ symmetry and leads to $\theta_{13}=0,\theta_{23}=\frac{\pi}{4}$. Moreover,
Eq. (\ref{relation}) then implies that sum of  elements in each of the rows of $M_{\nu f}$ is
equal. As discussed at several places \cite{reviews}, this implies  a $Z_2$ magic symmetry and
$M_{\nu f}$ in this limit is thus invariant under a $Z_2\times Z_2$ symmetry which implies TBM
mixing. Introducing imaginary part in $a,y$ can lead to the required deviation from the TBM mixing.
The following choice, 
\be \label{sol1} 
(a_1,~a_2,~x_1,~x_2)=(0.02369,~ 0.00512,~ 0.03159,~ -0.05003)~{\rm eV}\ee
minimizes the relevant $\chi^2$ and leads to the central values for $\theta_{12},~\theta_{13}$ and
the solar to atmospheric mass ratio. Overall scale is fixed from the value of the atmospheric mass
scale. The resulting masses are
\be \label{fit}
(m_{\nu_1},~m_{\nu_2},~m_{\nu_3})=(0.0497,~ 0.0505,~ 0.0008)~{\rm eV}~.\ee
This pattern corresponds to the inverted hierarchy. The atmospheric mixing angle and CP phase remain
maximal as they follow the relations, Eq. (\ref{predictions}) which are
independent of the choice of
parameters. The present data however favor non-maximal $\theta_{23}$ in the first quadrant. This
can be accommodated here by allowing complex $x_{1,2}$, $a_{1,2}$. A
possible but not unique choice giving correct
central values for all the mixing angles and $\Dms$, $\Dma$ is given by:
\be \label{sol2} 
(a_1,~a_2,~x_1,~x_2)=(-0.01327 + 0.06426 i,~ -0.00022,~ -0.04039,~ 0.08653)~{\rm eV}\ee

Before closing, let us note some interesting aspects of two contributions in $M_{\nu}$.
$M_{i\nu}M_{i\nu}^\dagger$ is diagonal and contains one massless and two degenerate states for each
$i=1,2$ when $x_{1,2},~a_{1,2}$ are real. Thus if only one of the contributions is present in
$M_\nu$ then the neutrino mixing matrix is given by $U_\omega^\dagger R$, where $R$ denotes a
rotation in the space of the degenerate eigenvalue by an arbitrary angle. The combined effect of
two contributions is to lift degeneracy and generate the solar scale and to give the correct mixing
angles. The massless state implied by each contribution is trimaximally mixed. The combined matrix
has a state with negligible mass and is almost bimaximally mixed, see Eqs. (\ref{sol1}, \ref{fit}).

\section{Summary}
\label{summary}
We explored in detail our recent proposal \cite{Joshipura:2013pga} of predicting a massless Majorana
neutrino and lepton mixing angles at the leading order by assuming that the neutrino and the charged
lepton symmetries appropriate for this are contained in some flavour group $G_f$. As argued, the DSG
of $U(3)$ are sufficient to provide at least one massless neutrino. Using available information on
such groups, we carried out detailed analysis for their viability to predict a massless neutrino and
suitable lepton mixing patterns at the leading order. The study encompassed very large number of
groups which include all DSG in class (X) defined earlier. As we discussed, their viability as
flavour symmetry and resulting mixing pattern can be worked out analytically. Main findings
applicable to groups in class (X) are:

\begin{itemize}
\item For all the groups in class (X), a  massless neutrino can remain
unmixed or it can be either
trimaximally  or bimaximally mixed. This result holds irrespective of
the order of the group
and is applicable to several known infinite group series. The above structures of the massless state
leads to incorrect solar mixing angle at the leading order.
\item The groups in this category can however lead  to a good zeroth order
prediction namely,
bimaximal for the mixing matrix in case of the inverted hierarchy which always predicts this 
pattern as shown analytically.
\item In case of the normal hierarchy, the different values of $\theta_{23}$
and $\theta_{13}$ can be
predicted depending on the group under consideration. The relation $\sin^2\theta_{12}\approx
\frac{2}{3}\cos^2\theta_{13}$ persists as explained above and the Dirac CP phase is trivial in all
such cases.
\end{itemize}

We further  numerically investigated 75 groups of order $<512$ tabulated by Ludl \cite{Ludl:2010bj}.
70 of these fall in category (X) and only the six of these are argued to qualify to be a symmetry
of the lepton sector. The mixing patterns implied by these groups for the normal hierarchy are
listed in Table \ref{NHresults}. All six can also lead to BM mixing with an inverted hierarchy.

The remaining 5 groups are also analyzed numerically and only one of them, namely the group
\group{432,239}, qualifies as a symmetry for a massless neutrino and  leads to predictions for all
three mixing angles fairly close to their best fit values in case of normal hierarchy. This group
thus provides a sole example of a flavour symmetry with massless neutrino and correct predictions
for all the mixing angles. A solution with inverted hierarchy is also found in this
case with the BM values of $\theta_{13}$ and $\theta_{23}$ but with $\sin^2\theta_{12}=0.21$.

Alternative ways to modify or avoid the unsatisfactory prediction of the solar angle are also
discussed briefly. These include relaxing assumption of leptons transforming as a 3-dimensional
irreducible representation and allowing more complicated residual symmetries. One can avoid wrong
prediction for the solar angle in the former case but it remains unpredicted. In this case, an
inverted hierarchy solution with \mt symmetry pattern can be obtained from the groups like
$\Sigma(2N^2)$ with $N \ge 3$ or even from their subgroups if $N$ is sufficiently large. In the
latter case, one can obtain all the mixing angles correctly as shown through two examples based
on the group \group{162,10} and $S_4(2)$.

\acknowledgments
A.S.J. thanks the Department of Science and Technology, Government of India for support under the
J. C. Bose National Fellowship programme, grant no. SR/S2/JCB-31/2010. K.M.P. thanks the Department
of Physics and Astronomy of the University of Padova for its support. He also acknowledges partial
support from the European Union network FP7 ITN INVISIBLES (Marie Curie Actions,
PITN-GA-2011-289442).

\end{document}